\documentclass[10pt]{elsart}
\usepackage{graphicx}
\usepackage{amssymb}

\def\dt{\Delta t}
\def\dx{\Delta x}
\def\p{\partial}
\def\grad{\nabla}

\def\d{\Delta}

\begin{document}

%%%%%%%%%%%%%%%%%%%%%%%%%%%%%%%%5%%%%%%%%%%%%%%%%%%%%%%%%%%%%%%
\begin{frontmatter}



\title{Error analysis in unconditionally stable coarsening algorithms}
\author{Mowei Cheng}
\ead{mowei.cheng@gmail.com}
\address{Metallurgy Division and Center for Theoretical and Computational Materials
Science, National Institute of Standards and Technology, Gaithersburg, Maryland 20899, USA}

%%%%%%%%%%%%%%%%%%%%%%%%%%%%%%%%%%%%%%%%%%%%
\begin{abstract}
% Text of abstract

In order to quantitatively study the accuracy of the unconditionally stable
coarsening algorithms, we calculate the Fourier space multi step error on the
order parameter field by explicitly distinguishing the analytic time $\tau$
and the algorithmic time $t$. The calculation determines the error in the order
parameter and the scaled correlations. This error contributes a correction term
in the analytic time step, which is crucial in understanding the accuracy in
unconditionally stable coarsening algorithms.

\end{abstract}

%%%%%%%%%%%%%%%%%%%%%%%%%%%%%%%%%%%%%%%%%%%%
\begin{keyword}
Unconditionally stable \sep Error analysis \sep Algorithmic time step \sep Multi step error \sep Coarsening

\PACS 05.10.-a \sep 02.60.Cb \sep 64.75.+g \sep 81.15.Aa
\end{keyword}

%%%%%%%%%%%%%%%%%%%%%%%%%%%%%%%%%%%%%%%%%%%%

\end{frontmatter}
%%%%%%%%%%%%%%%%%%%%%%%%%%%%%%%%%%%%%%%%%%%%%%%%%%%%%%%%%%%%%%%%%%%%%%

% main text
%\section{}
%\label{}

%%%%%%%%%%%%%%%%%%%%%%%%%%%%%%%%%%%%
\section{Introduction}

The Partial Differential Equations (PDEs) that governs conserved and
non-conserved scalar coarsening systems are the Cahn-Hilliard (CH) \cite{Cahn1} and
Allen-Cahn (AC) equations \cite{Allen1} respectively.
Examples are found in polymer mixtures \cite{Wiltzius1}, alloys
\cite{Shannon1,Gaulin1}, liquid-crystals \cite{Mason1,Chuang1}, and in
cosmology \cite{Laguna1}. These equations,
as well as their various extended forms, are an active area of research.
They model the phase separation that occurs after a quench from a high temperature
disordered phase into two distinct phases at low temperatures. The pattern
of the two phase regions coarsens as the time increases, i.e., the length-scale
of these regions grows. The scaled correlation is an important probe to characterize
the late time scaling regime, where the dynamics are dominated by a
single length scale, the average domain size $L$, which increases with
a power law in time $\tau$, $L(\tau) \sim \tau^{\alpha}$ \cite{Bray1}.
For the scalar order parameter systems considered in this paper,
$\alpha=1/3$ and $1/2$ for conserved and and non-conserved systems,
respectively \cite{Bray1}. To study the coarsening systems, one
needs to choose a random initial condition, which corresponds to a
high temperature disordered phase. Since it is essentially not possible
to find an exact solution for random initial conditions, computer
simulation plays an important role in understanding these systems.

Because of the nature of slow coarsening process in the late time scaling regime, the main
challenge in computer simulation of coarsening systems is to develop an efficient
algorithm. The most straightforward approach is the Euler algorithm, which must
employ a time step $\dt_{Eu}\sim (\Delta x)^4$ and $(\Delta x)^2$ to maintain stability
for conserved and non-conserved systems, respectively, where $\dx$ is the lattice
spacing. On the other hand, one has to use a lattice spacing $\dx < \xi$ to
resolve the interfacial profile, where $\xi$ is the interfacial width.
This implies that the Euler update is inefficient, and in practice
computationally costly to use to evolve large systems until the late
time scaling regime. Various computational algorithms have been developed
by increasing the algorithmic time step $\dt$ compared to the simplest Euler
discretization. For example,
the Cell Dynamical System (CDS) \cite{Oono1,Oono2,Puri1} was broadly
applied \cite{Shinozaki1,Zapotocky1} as an efficient technique.
However, a transformation of CDS to PDE \cite{Teixeira1,Oono3} showed that CDS
changes the form of the equation of motion within the same universality class
and gets a less strict stability threshold for its Euler update, but the Euler
time step is still limited to $(\Delta x)^4$ and
$(\Delta x)^2$ for conserved and non-conserved systems, respectively.
More recently, conditionally stable algorithms such as Fourier spectral
methods \cite{Chen1,Zhu1} have been developed by introducing implicit terms in the
update equation, and have been shown to allow an increase of the algorithmic time
step $\dt$ by two orders of magnitude. However, none of these methods can
adjust to the naturally slowing dynamics - they eventually become more and
more inefficient at the late time scaling regime.

The recently developed unconditionally stable algorithms \cite{Eyre1,Eyre2,Vollmayr1}
overcomes the difficulty of a fixed time step for maintaining stability and allows for arbitrarily
large algorithmic time steps. In CH equation, the algorithm allows an optimal driving scheme
$\dt=A \tau^{2/3}$ with the error in correlations scales as $\sqrt{A}$ for small prefactor
$A$ \cite{Cheng1,Cheng2}. This implies that any driving scheme that is
slower than the optimal one is wastefully accurate because the error is negligible
in correlations \cite{Cheng2}. This algorithm has also been applied to other
interesting systems, such as the Phase Field Crystal (PFC)
model \cite{Elder1,Elder2} and the Swift-Hohenberg (SH) equation
and showed significantly improved efficiency \cite{Cheng3}.

While we have no doubt about the efficiency of unconditionally stable algorithms, the
accuracy of unconditionally stable algorithm, remains somewhat mysterious, and some open
questions remain unanswered. For example, why does the single step error accumulate
over time \cite{Cheng1,Cheng2}? why does the fitting for
the maximal analytic time step in non-conserved system get worse as $\tilde{a}$
[see Eq. (\ref{eq:tilde-a}) for a precise definition] increases? In this paper, by explicitly
calculating the Fourier space multi step error using Eq. (\ref{eq:multi_step_error}),
we are able to answer these questions. The rest of the paper is
organized as follows. In Section 2, we briefly review the results in the previous
work \cite{Eyre1,Eyre2,Vollmayr1,Cheng1,Cheng2} and introduces the basic
concepts and notations of unconditionally stable algorithms. In Section 3,
we study the Fourier space multi step error, where the distinction of the analytic time and the
algorithmic time is necessary. Based on this result, in Section 4, we determine
the accuracy in the scaled structure factor $g(x)$ and $h(x)$. We then calculate the
analytic time step in Section 5, taking into
account the error in $h(x)$. The summary are presented in Section 6. In the
appendix, we derive the scaling of field derivatives for all modes in Fourier space. For simplicity
but without loss of generality, we restrict our analysis to two dimensions (2D).

%%%%%%%%%%%%%%%%%%%%%%%%%%%%%%%%%%%%%%%
\section{Review of Unconditionally stable algorithms}

For coarsening systems, a suitable free energy functional to describe the ordered phase is \cite{Bray1}
\begin{equation}
F[\phi]=\int d^2x \left[\frac{1}{2} |\nabla \phi|^2 + V(\phi) \right],
        \label{eq:TDGLbasic}
\end{equation}
where $\phi({\bf x},t)$ is the order parameter, and the potential $V(\phi)$
has a double-well structure $V(\phi)=(\phi^2-1)^2/4$. The two
minima of $V$ correspond to the two equilibrium phases $\phi=\pm 1$. The term $V(\phi)$ describes
the energy in the bulk. The gradient-squared term in Eq. (\ref{eq:TDGLbasic}) associates
an energy cost to an interface between the two phases.

When the order parameter is conserved under the dynamics, the equation of motion
can be written in the form of a continuity equation,
$\partial \phi/\partial t= -\nabla \cdot {\bf j}$, with current
${\bf j} = - M \nabla (\delta F/\delta \phi)$, where $M$ is the mobility.
Absorbing $M$ into the time scale, we obtain the dimensionless form of
Cahn-Hilliard (CH) equation \cite{Cahn1}:
\begin{eqnarray}
\frac {\partial \phi} {\partial \tau} = \nabla^2 \frac{\delta F}{\delta \phi}
= -\nabla^2(\nabla^2 \phi+\phi-\phi^3).
        \label{eq:CHbasic}
\end{eqnarray}
In the case where the order parameter is not conserved by the dynamics,
the simplest dissipative equation for the time evolution of the field $\phi$ is
\begin{eqnarray}
\frac {\partial \phi} {\partial \tau} = -\frac{\delta F}{\delta \phi} = \nabla^2 \phi+\phi-\phi^3.
        \label{eq:ACbasic}
\end{eqnarray}
Eq. (\ref{eq:ACbasic}) describes that the rate of change of $\phi$ is proportional
to the gradient of the free-energy functional in functional space, and is often
called the Allen-Cahn (AC) equation \cite{Allen1}.

In order to obtain unconditionally stable algorithms for these equations, we introduce a
mix of explicit and implicit terms in the update \cite{Eyre1,Eyre2,Vollmayr1}. For the CH equation, we have
\begin{eqnarray}
\phi_{t+\dt} - (a_1 -1) \dt \nabla^2 \phi_{t+\dt}
- (a_2-1) \dt \grad^4 \phi_{t+\dt} \nonumber \\
= \phi_t - \dt (a_1 \nabla^2 \phi_t + a_2 \grad^4 \phi_t - \nabla^2 \phi_t^3),
        \label{eq:CHdirect}
\end{eqnarray}
and for the AC equation, we have
\begin{eqnarray}
\phi_{t+\dt} + (a_1 -1) \dt \phi_{t+\dt}
+ (a_2-1) \dt \grad^2 \phi_{t+\dt} \nonumber \\
= \phi_t + \dt (a_1 \phi_t + a_2 \grad^2\phi_t - \phi_t^3).
        \label{eq:ACdirect}
\end{eqnarray}
Implicit terms ($\phi_{t+\dt}$) are on the left and the explicit terms ($\phi_t$)
are on the right. Unconditionally stable algorithms are obtained for $a_1>2$ and $a_2<1/2$ \cite{Eyre1,Eyre2,Vollmayr1}.
In Fourier space, Eq. (\ref{eq:CHdirect}) can be directly solved as
\begin{equation}
\phi_k(t+\dt)= \phi_k(t) + \dt_{eff}(k,\dt) \frac{\p {\phi}_k}{\p \tau},
        \label{eq:dteff}
\end{equation}
where the $k$-dependent effective time step is
\begin{eqnarray}
\dt_{eff}(k,\dt) \equiv \frac {\dt}{1 + \dt k^2 [(a_1-1) + (1-a_2) k^2]},
        \label{eq:CHdteff}
\end{eqnarray}
and $\p {\phi}_k/\p \tau$ is the Fourier transform of $\p \phi/\p \tau$ in
Eq. (\ref{eq:CHbasic}) and is a function of ${\phi}_k(t)$. As
Eq. (\ref{eq:CHdteff}) reveals, the unconditionally stable algorithm
has a mode-dependent effective time step $\dt_{eff}(k,\dt)$. The optimal driving scheme
for CH equation is an algorithmic time step $\dt=A\tau^{2/3}$,
where the prefactor $A$ determines the accuracy \cite{Cheng1,Cheng2}.

On the other hand, the effective time-step for AC equation is
\begin{eqnarray}
\dt_{eff}(k,\dt) \equiv \frac {\dt}{1 + \dt [(a_1-1) + (1-a_2) k^2]},
        \label{eq:ACdteff}
\end{eqnarray}
which indicates that
\begin{eqnarray}
\dt_{eff}(k,\dt) \le \dt_{eff}(k,\infty) < \frac{1}{a_1-1},
        \label{eq:ACdteff_maximum}
\end{eqnarray}
where $a_1-1>0$ and $1-a_2>0$ are used. We cannot obtain an accelerated algorithm
in non-conserved systems since there is an upper bound for the effective time-step
$\dt_{eff}(k,\dt)$. In fact, the optimal driving scheme for AC equation is an algorithmic time
step $\dt = \infty$ \cite{Cheng1}.

%%%%%%%%%%%%%%%%%%%%%%%%%%%%%%%%%%%%%%%
\section{Fourier space multi step error}

In a computer simulation, assuming that we have updated the system $m-1$ times, the
algorithmic time is $t_{m-1}$, and the field is $\phi_k(t_{m-1})$ in Fourier space.
We next update the system using an algorithmic time step $\dt_m$, and the
field evolves according to Eq. (\ref{eq:dteff}):
\begin{equation}
\phi_k(t_m)= \phi_k(t_{m-1}) + \dt_{eff}(k,\dt_m) \frac{\p {\phi}_k}{\p \tau},
\end{equation}
where $t_m \equiv t_{m-1}+\dt_m$. To obtain the error of this unconditionally
stable algorithm $\d \phi_k$, we need to compare the field evolved by an unconditionally stable
algorithm $\phi_k(t)$ to the exact dynamics $\phi_k(\tau)$ evolved by the same
amount of energy \cite{Cheng1,Cheng2}. The analytic time step $\d \tau$ and the analytic
time $\tau$ can be determined by the energy evolution of each update \cite{Cheng2}.
At an analytic time $\tau_{m-1}$, the system evolution after an analytic time step $\d \tau_m$ is
governed by the Taylor expansion:
\begin{equation}
\phi_k(\tau_m)=\phi_k(\tau_{m-1}) + \sum_{n=1}^\infty
\frac{\p^n \phi_k}{\p \tau^n} \frac{\d \tau_m^n}{n!},
        \label{eq:Taylor}
\end{equation}
where $\tau_m \equiv \tau_{m-1}+\d \tau_m$. The Fourier space multi step error up to the $m$th update is
\begin{eqnarray}
\d \phi_k (m) \equiv \phi_k(t_{m})-\phi_k(\tau_{m}).
\end{eqnarray}
Note that
\begin{eqnarray}
\d \phi_k (m) = \d \phi_k (m-1) + \delta \phi_k (m),
\end{eqnarray}
where
\begin{eqnarray}
\delta \phi_k (m) = \left[ \dt_{eff}(k,\dt_m) - \d \tau_m \right] \frac{\p \phi_k}{\p \tau}
-\sum_{n=2}^\infty \frac{\p^n \phi_k}{\p \tau^n} \frac{\d \tau_m^n}{n!}.
\end{eqnarray}
We obtain that
\begin{eqnarray}
\d \phi_k (m)=\sum_m \delta \phi_k (m) \approx \int \delta \phi_k dm,
        \label{eq:multi_step_error}
\end{eqnarray}
where we use the integral to approximate the summation. Eq. (\ref{eq:multi_step_error}) explains
the numerical results that the single step error
$\delta \phi_k (m)$ accumulates over time, which was observed in previous studies \cite{Cheng1,Cheng2}.
Note that as long as there is a distinction between the analytic time $\tau$ (time step
$\d \tau$) and the algorithmic time $t$ (time step $\dt$), we can use Eq. (\ref{eq:multi_step_error})
to calculate the Fourier space multi step error.

As an interesting application of the above result Eq. (\ref{eq:multi_step_error}), we study the accuracy in
the simplest Euler algorithm. The Euler algorithm has a mode-independent fixed time step $\dt_{Eu}$ to
update the system, and $\dt_{eff}(k,\dt)=\dt_{Eu}$. In conserved systems, evolving from $\tau_0$ to $\tau_{max}$
with a fixed time step $\dt_{Eu} \approx \d \tau \sim d \tau/dm$ \cite{Cheng2}, we
obtain the Fourier space multi step error as $kL \approx 1$:
\begin{eqnarray}
\d \phi_k
&\sim& - \int_{\tau_0}^{\tau_{max}} \frac{1}{\dt_{Eu}} \frac{\dt_{Eu}^2}{2} \tau^{-5/3} d\tau
\nonumber \\
&=& \frac{3 \dt_{Eu}}{4} \left(\tau_{max}^{-2/3} - \tau_0^{-2/3} \right)
\approx -\frac{3 \dt_{Eu}}{4} \tau_0^{-2/3},
\end{eqnarray}
where we ignore the higher order terms in the sum and
$\p^2 \phi_k/\p \tau^2 \sim \tau^{-5/3}$ as $k \approx 1/L$ is used \cite{Cheng2}.
We can use this to determine the error in the scaled structure factor
$g(kL)=\langle \phi_k \phi_{-k} \rangle/L^2$, where the angle brackets indicate an average
over orientations and initial conditions. The absolute error is
\begin{eqnarray}
\d g(kL) \approx \left| \frac{2 \d \phi_k \phi_k}{L^2} \right|
\sim \frac{\dt_{Eu} \tau_0^{-2/3}}{L},
	\label{eq:structureg_CH_Euler}
\end{eqnarray}
where $\phi_k \sim L$ as $k \approx 1/L$ is used \cite{Cheng2}. Since
$L \sim \tau^{1/3} \to \infty$ at late times, the error $\d g(kL) \to 0$ at late times.
This is why the Euler algorithm --- or any fixed time step algorithm --- is wastefully accurate.

%%%%%%%%%%%%%%%%%%%%%%%%%%%%%%%%%%%%%%%
\section{Error in scaled structures}

In this Section, we calculate the error in scaled structures due to the Fourier space multi step
error determined in Section 3 with the optimal driving schemes discussed in Section 2.

%%%%%%%%%%%%%%%%%%%%%%%%%%%%
\subsection{Conserved systems}

\begin{figure}[htb]
\begin{center}
\includegraphics[width=12cm]{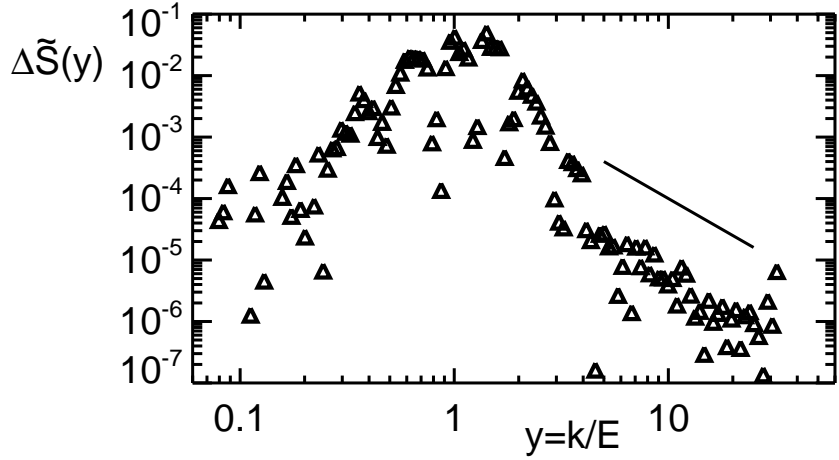}
\caption{The absolute difference between unconditionally stable and Euler updates,
$\d \tilde{S}(y)$ vs. $y \equiv k/E$ for system size $L_\infty =512$ at $\tau=1024$.
The Euler update used $\dt_{Eu} = 0.03$ and the unconditionally stable update
used $\dt= 0.01 \tau^{2/3}$. The plot shows that $\d \tilde{S}(y) \sim y^{-2}$ as $y \gtrsim 1$,
in agreement with Eq. (\ref{eq:structureg_CH}).}
        \label{FIG:error_in_structure}
\end{center}
\end{figure}

Using the formula discussed in Section 3, we obtain the Fourier space single
step error with the optimal driving scheme $\dt=A\tau^{2/3}$ \cite{Cheng2}
\begin{eqnarray}
\delta\phi_k \sim \frac{L_0 C \sqrt{a_1-1}}{12 B^{1/3}} A^{3/2} \tau^{-1/6} k^{-1/2}
        \label{eq:single_step_error_CH}
\end{eqnarray}
where $A$, $B$ and $C$ are constants \cite{Cheng2}, and
$\p \phi_k/\p \tau \sim \tau^{-5/6} k^{-1/2}$ as $k \gtrsim 1/L$ is used [see Appendix].
For a small $A$, evolving from $\tau_0$ to $\tau_{max}$ with time-step
$\dt = A \tau^{2/3} \approx \d \tau \sim d\tau/dm$, we obtain the Fourier space
multi step error:
\begin{eqnarray}
\d \phi_k
&\sim& \frac{L_0 C \sqrt{a_1-1}}{12 B^{1/3}} k^{-1/2}
\int_{\tau_0}^{\tau_{max}} \frac{1}{A\tau^{2/3}} A^{3/2} \tau^{-1/6} d\tau
\nonumber \\
&\approx& \frac{L_0 C \sqrt{a_1-1}}{2 B^{1/3}} \sqrt{A}k^{-1/2} \tau_{max}^{1/6}.
\end{eqnarray}
We can use this result to determine the error in the scaled structure factor
$g(kL)=\langle \phi_k \phi_{-k} \rangle/L^2$, where the angle brackets indicate an average
over orientations and initial conditions. The absolute error is
\begin{eqnarray}
\d g(kL) \approx \left| \frac{2 \d \phi_k \phi_k}{L^2} \right|
\sim \frac{L_0 C \sqrt{a_1-1}}{B^{1/3}}\frac{\sqrt{A}}{(kL)^2},
        \label{eq:structureg_CH}
\end{eqnarray}
where $\phi_k \sim \tau_{max}^{-1/6}k^{-3/2}$ as $k \gtrsim 1/L$ is used [see Appendix].
We can subsequently obtain the absolute error in the scaled derivative structure factor
$h(kL)=(kL)^2g(kL)$:
\begin{eqnarray}
\d h(kL) = (kL)^2 \d g(kL) \sim \frac{L_0 C \sqrt{a_1-1}}{B^{1/3}}\sqrt{A}.
        \label{eq:structureh_CH}
\end{eqnarray}
To test Eq. (\ref{eq:structureg_CH}), we measure $S(k,t) = \langle \phi_k \phi_{-k} \rangle$,
and obtain the time-independent scaling function $\tilde{S}(y) \equiv E^2 S(y E)$ using the
energy density $E$, where $y \equiv k/E$.
In Fig.~\ref{FIG:error_in_structure} we plot the absolute difference between scaled
structures obtained using an Euler algorithm and an unconditionally stable algorithm,
$\d \tilde{S}(y)$ vs. $y$. We find that $\d \tilde{S}(y) \sim y^{-2}$ as $y \gtrsim 1$,
which implies $\d g(x) \sim x^{-2}$ as $x \gtrsim 1$, where $x \equiv kL$. We confirm
the prediction in Eq. (\ref{eq:structureg_CH}).

%%%%%%%%%%%%%%%%%%%%%%%%%%%%
\subsection{Non-conserved systems}

Using the formula discussed in Section 3, we obtain the Fourier space single
step error with the maximal driving scheme $\dt=\infty$ \cite{Cheng1}
\begin{eqnarray}
\delta \phi_k \sim \left[\frac{1}{a_1-1} - \d \tau_{\infty}\right] \tau^{-3/4} k^{-1/2},
        \label{eq:single_step_error_AC}
\end{eqnarray}
where $\p \phi_k/\p \tau \sim \tau^{-3/4} k^{-1/2}$ as $k \gtrsim 1/L$ is used [see Appendix].
Evolving from $\tau_0$ to $\tau_{max}$ with time-step $\d \tau_{\infty} \sim d\tau/dm$,
we obtain the Fourier space multi-step error:
\begin{eqnarray}
\d \phi_k
&\sim& k^{-1/2} \int_{\tau_0}^{\tau_{max}} \frac{1}{\d \tau_{\infty}}
\left[\frac{1}{a_1-1}-\d \tau_{\infty}\right] \tau^{-3/4} d\tau
\nonumber \\
&\approx& k^{-1/2} \tau_{max}^{1/4}
4 \left[\frac{\tilde{a}}{\xi \tan^{-1}\left(\tilde{a}/\xi\right)} - 1 \right],
\end{eqnarray}
where
\begin{eqnarray}
\tilde{a} \equiv \sqrt{\frac{1-a_2}{a_1-1}}.
	\label{eq:tilde-a}
\end{eqnarray}
We then obtain the absolute error in scaled structure:
\begin{eqnarray}
\d g(kL) \approx \left| \frac{2 \d \phi_k \phi_k}{L^2} \right|
\sim \frac{8}{(kL)^2} \left[\frac{\tilde{a}}{\xi \tan^{-1}\left(\tilde{a}/\xi\right)} - 1 \right],
        \label{eq:structureg_AC}
\end{eqnarray}
where $\phi_k \sim \tau_{max}^{-1/4}k^{-3/2}$ as $k \gtrsim 1/L$ is used [see Appendix].
We can subsequently obtain the absolute error in the scaled derivative structure factor
$h(kL)=(kL)^2g(kL)$:
\begin{eqnarray}
\d h(kL) = (kL)^2 \d g(kL)
\sim 8\left[\frac{\tilde{a}}{\xi \tan^{-1}\left(\tilde{a}/\xi\right)} - 1 \right].
	\label{eq:structureh_AC}
\end{eqnarray}
This result indicates that the multi-step error in correlations increases
with $\tilde{a}$.

%%%%%%%%%%%%%%%%%%%%%%%%%%%%%%%%%%%%%%%%
\section{Error in analytic time step}

In this Section, we calculate the error in analytic time step due to the error in
scaled derivative structures ($\d h$) calculated in Section 4.

%%%%%%%%%%%%%%%%%%%%%%%%%%%%
\subsection{Conserved systems}

CH systems are purely dissipative systems --- the energy density $E$
monotonically decreases with the analytic time $\tau$ with the relation
$E \propto 1/L \propto \tau^{-1/3}$ \cite{Bray1}. We can calculate the
analytic time in terms of the energy density $E$: $\tau = B_{CH}/E^3$, where
the prefactor $B_{CH}$ can be numerically determined by requiring $\d \tau=\dt$
as $\dt \to 0$ in the late-time scaling regime since unconditionally
stable algorithms are arbitrarily accurate as $\dt \to 0$.
We can then calculate the analytic time step by differentiating $\tau$
with respect to $E$:
\begin{equation}
\d \tau=\frac{-3B_{CH} \Delta E}{E^4}=\frac{-3 \Delta E \tau^{4/3}}{B_{CH}^{1/3}},
	\label{eq:CHdts}
\end{equation}
and $\Delta E$ can be calculated by integrating from each Fourier mode:
\begin{eqnarray}
\Delta E
&=& \int_0^{1/\xi} d^2k \frac{1}{(2\pi)^2}
\left\langle \left(\frac{\delta F}{\delta \phi_k}\right) [\phi_k(t+\dt)-\phi_k(t)]
\right\rangle
\nonumber \\
&=&-\int_0^{1/\xi} d^2k \frac{1}{(2\pi k)^2} \dt_{eff}(k,\dt) T_k,
        \label{eq:CHde}
\end{eqnarray}
where the time derivative $\p \phi_{-k}/\p \tau=-k^2 \delta F/\delta \phi_k$ from
Eq. (\ref{eq:CHbasic}), and $\phi_k(t+\dt)-\phi_k(t)=\dt_{eff} \p \phi_k/\p \tau$
from Eq. (\ref{eq:dteff}) are used. The time-derivative correlation function
$T_k$ \cite{Bray1,Bray2,Rutenberg1} has a natural scaling form given by
\begin{eqnarray}
T_k \equiv \left \langle \frac{\p \phi_k}{\p \tau}
\frac{\p \phi_{-k}}{\p \tau} \right \rangle
= \left( \frac{dL}{d \tau} \right)^2 h(kL)
= \frac{L_0^2h(kL)}{9 \tau^{4/3}},
\end{eqnarray}
where $L=L_0\tau^{1/3}$,  $h(x)$ is the $2$D scaling function
\cite{Rutenberg1}, and $L_0$ is a constant. Taking the error in scaled structures
into account, $h(x)$ takes the form \cite{Rutenberg1}
\begin{eqnarray}
h(x) = \frac{C_{CH}}{x} \pm \d h,
\end{eqnarray}
as $x \gtrsim 1$, where $C_{CH}$ is a constant and $\d h \ge 0$ is the absolute
error in Eq. (\ref{eq:structureh_CH}). The sign $+$ or $-$ will be determined later.
Using the optimal driving scheme $\dt=A \tau^{2/3}$ \cite{Cheng2},
we can solve for $\Delta E$ in Eq. (\ref{eq:CHde}) and for the analytic
time step $\d \tau$ in Eq. (\ref{eq:CHdts}). We have
\begin{eqnarray}
\d \tau
&=& \frac{C_{CH}L_0^2 \dt}{6\pi B_{CH}^{1/3}} \int_0^{L/\xi} \frac{dx}{x^2[1+(a_1-1)Ax^2/L_0^2]}
\nonumber \\
&\pm& \frac{\d h L_0^2 \dt}{6\pi B_{CH}^{1/3}} \int_{x_0}^{L/\xi}\frac{dx}{x[1+(a_1-1)Ax^2/L_0^2]},
        \label{eq:CH-dts-dt}
\end{eqnarray}
Solving the integral, we obtain
\begin{eqnarray}
\d \tau=\dt \left[1 - \left(\frac{L_0 C_{CH} \sqrt{a_1-1}}{12 B_{CH}^{1/3}}+d\right) \sqrt{A}
+\mathcal{O}(A) \right],
\end{eqnarray}
where
\begin{eqnarray}
d\sim \frac{CL_0^3\sqrt{a_1-1}}{12\pi B_{CH}^{2/3}} \ln \left[\frac{L_0^2}{(a_1-1)A}\right],
\end{eqnarray}
and Eq. (\ref{eq:structureh_CH}) is used. Note that we drop the $+$ sign in Eq. (\ref{eq:CH-dts-dt})
according to the requirement
of $\d \tau \le \dt$ implied by Eq. (\ref{eq:CHdteff}). We have $1-\d \tau/\dt \sim \sqrt{A}$,
which is consistent with the previous results \cite{Cheng1,Cheng2}.

%%%%%%%%%%%%%%%%%%%%%%%%%%%%
\subsection{Non-conserved systems}

Similar to the calculation in conserved systems, we can calculate the analytic time step by
differentiating $\tau$ with respect to $E$:
\begin{eqnarray}
\d \tau = \frac{-2B_{AC} \Delta E}{E^3}= \frac{-2 \Delta E \tau^{3/2}}{B_{AC}^{1/2}}.
        \label{eq:ACdts}
\end{eqnarray}
Integrating $\d E$ from each Fourier mode, we have
\begin{eqnarray}
\Delta E
&\approx& \int_0^{1/\xi} d^2k \frac{1}{(2\pi)^2}
\left\langle \left(\frac{\delta F}{\delta \phi_k}\right) [\phi_k(t+\dt)-\phi_k(t)] \right\rangle
\nonumber \\
&=&-\int_0^{1/\xi} d^2k \frac{1}{(2\pi)^2} \dt_{eff}(k,\dt) T_k,
        \label{eq:ACde}
\end{eqnarray}
where the time derivative $\p \phi_{-k}/\p \tau=- \delta F/\delta \phi_k$ from Eq. (\ref{eq:ACbasic}), and
$\phi_k(t+\dt)-\phi_k(t)=\dt_{eff} \p \phi_k/\p \tau$ from Eq. (\ref{eq:dteff}) are used. The time-derivative correlation function
$T_k$ \cite{Bray1,Bray2,Rutenberg1} has a natural scaling form of
\begin{eqnarray}
T_k \equiv
\left \langle \frac{\p \phi_k}{\p \tau}
\frac{\p \phi_{-k}}{\p \tau} \right \rangle
= \left( \frac{dL}{d \tau} \right)^2 h(kL)
= \frac{L_0^2 h(kL)}{4 \tau},
        \label{eq:ACT(k)scaling}
\end{eqnarray}
where $L=L_0 \tau^{1/2}$, $h(x)$  is the 2D scaling function \cite{Rutenberg1},
and $L_0$ is a constant. Taking the error in scaled structures
into account, $h(x)$ takes the form \cite{Rutenberg1}
\begin{eqnarray}
h(x) = \frac{C_{AC}}{x} \pm \d h,
\end{eqnarray}
as $x \gtrsim 1$, where $C_{AC}$ is a constant and $\d h \ge 0$ is the absolute
error in Eq. (\ref{eq:structureh_AC}). The sign $+$ or $-$ will be determined
later. We can solve for $\Delta E$ in Eq. (\ref{eq:ACde}) and for $\d \tau$ from
Eq. (\ref{eq:ACdts}). We have
\begin{eqnarray}
\d \tau
&=&\frac{C_{AC} \dt L_0}{4\pi B_{AC}^{1/2}} \frac{\tan^{-1}\left(\sqrt{\frac{\dt(1-a_2)}{1+\dt(a_1-1)}}\frac{1}{\xi}\right)}{\sqrt{[1+\dt(a_1-1)]\dt(1-a_2)}}
\nonumber \\
&\pm& \frac{\d h L_0^2 \tau^{1/2}}{8\pi B_{AC}^{1/2} (1-a_2)}
\ln \left( 1+\frac{\dt(1-a_2)}{[1+\dt(a_1-1)] \xi^2} \right).
        \label{eq:dtsNC}
\end{eqnarray}
As $\dt = \infty$, we obtain the maximal analytic time step:
\begin{eqnarray}
\d \tau_{\infty}
=\frac{\xi \tan^{-1}(\tilde{a}/\xi)}{\sqrt{(a_1-1)(1-a_2)}}
-\frac{L_0^2 \ln (1+\tilde{a}^2/\xi^2)\d h \tau^{1/2}}{8\pi B_{AC}^{1/2}(1-a_2)},
\end{eqnarray}
where Eq. (\ref{eq:structureh_AC}) is used. Note that we drop the $+$ sign in Eq. (\ref{eq:dtsNC})
according to the requirement of $\d \tau_{\infty} < 1/(a_1-1)$ implied by
Eq. (\ref{eq:ACdteff_maximum}). Define
\begin{eqnarray}
\widetilde{\d \tau}_{\infty}
\equiv \d \tau_{\infty} \sqrt{(a_1-1)(1-a_2)}
=\xi\tan^{-1}\left(\frac{\tilde{a}}{\xi}\right)-f(\tilde{a}) \tau^{1/2},
	\label{eq:widetilde_tau}
\end{eqnarray}
where
\begin{eqnarray}
f(\tilde{a}) \sim \frac{L_0^2}{\pi B_{AC}^{1/2}}
\left[\frac{1}{\xi \tan^{-1}\left(\tilde{a}/\xi\right)} - \frac{1}{\tilde{a}} \right]
\ln \left(1+\frac{\tilde{a}^2}{\xi^2}\right).
    \label{eq:f(a)}
\end{eqnarray}

\begin{figure}[htb]
\begin{center}
\includegraphics[width=12cm]{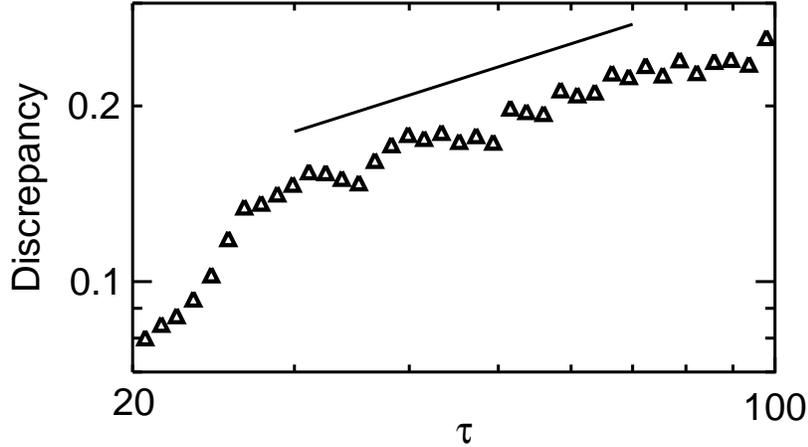}
\caption{The discrepancy in the maximal analytic time step
$f(\tilde{a}) \tau^{1/2}$ vs. the analytic time $\tau$ in the
scaling regime ($20 \le \tau \le 100$) with $\tilde{a}=2$ ($a_1=3$ and $a_2=-7$) for system size $L_\infty=512$.
$\xi$ is taken to be $0.92$ for best fit by eye. The discrepancy shows a $\tau^{1/2}$ behavior, as predicted in
Eq. (\ref{eq:widetilde_tau}).}
        \label{FIG:dtau}
\end{center}
\end{figure}

In a previous paper \cite{Cheng1}, we fitted the time-averaged $\widetilde{\d \tau}_{\infty}$ with $\xi\tan^{-1}(\tilde{a}/\xi)$, and
found that the fitting gets worse as $\tilde{a}$ increases. This is consistent with the fact that
$f(\tilde{a})$ monotonically increases with $\tilde{a}$. In addition, for a chosen $\tilde{a}$, in
Fig. \ref{FIG:dtau}, we test the $\tau$-dependence of the
discrepancy $f(\tilde{a}) \tau^{1/2}$ in the scaling regime and find excellent agreement.

%%%%%%%%%%%%%%%%%%%%%%%%%%%%%%%%%%%%%%%
\section{Summary}

We find it is interesting to compare the analytic time $\tau$ with the algorithmic time $t$.
The analytic time $\tau$, determined by energy evolution, is a more fundamental quantity.
The unconditionally stable update Eq. (\ref{eq:dteff}) is directly related to
$\p \phi_k/\p \tau$, and all scaling analysis is in analytic time $\tau$. The algorithmic time
step $\dt$, on the other hand, is a more practical quantity. It appears in the driving
scheme and is used to express the Fourier space effective time step and to calculate
the analytic time step. Our accuracy study on the scaled correlations $g(x)$ and $h(x)$ and subsequently
on the analytic time step $\d \tau$ based on the error accumulation relation
Eq. (\ref{eq:multi_step_error}) were verified by numerical simulations. The
analytic time step, the effective time step, the algorithmic time step, and the Fourier
space multi step error are fundamental in understanding the error behavior in unconditionally
stable algorithms for coarsening systems. The methodology can be readily generalized
to study other systems. We hope to report that work in a subsequent paper.

%%%%%%%%%%%%%%%%%%%%%%%%%%%%%%%%%
\section{Acknowledgments}

I thank James Warren for useful discussions.

\appendix
% \section{}
% \label{}

%%%%%%%%%%%%%%%%%%%%%%%%%%%%%%%%%%%%%%%%%%%%%%%%%%%%%%
\section{Scaling of field derivatives in Fourier space}

In order to explore the accuracy of unconditionally stable algorithms in Fourier space,
it is necessary to know the scaling of field derivatives for all modes.
The structure factor $S(k) = \langle|\phi_k|^2\rangle =L^2g(kL)$, where
$g(kL) \sim (kL)^{-3}$ as $k \gtrsim 1/L$ \cite{Bray1}. Therefore we obtain
\begin{eqnarray}
\phi_k \sim L (kL)^{-3/2} = k^{-3/2} L^{-1/2}.
\end{eqnarray}
Previous studies \cite{Bray1,Rutenberg1} showed that
$\p \phi_k/\p \tau = (dL/d \tau) k \phi_k$ as $kL \gtrsim 1$. Then we obtain the
time-derivative correlation function
\begin{eqnarray}
T(k)=\langle|\p \phi_k/\p \tau|^2\rangle=(dL/d \tau)^2k^2\langle|\phi_k|^2\rangle=(dL/d \tau)^2h_1(kL),
\end{eqnarray}
where the scaling function $h_1(kL)=k^2L^2g(kL) \sim (kL)^{-1} \sim L^{-1}$ as $k \gtrsim 1/L$.
Subsequently we obtain
\begin{eqnarray}
\frac{\p \phi_k}{\p \tau} \sim \frac{dL}{d \tau} L^{-1/2} k^{-1/2}.
\end{eqnarray}
The generalization of higher order time-derivative correlations is
$\langle|\p^n \phi_k/\p \tau^n |^2\rangle
\sim (dL/d \tau)^2k^2\langle|\p^{n-1} \phi_k/\p \tau^{n-1} |^2\rangle$,
where ``$\sim$'' indicates that generally the left hand side may not exactly
equal to the right hand side. Applying this relation yields
\begin{eqnarray}
\langle|\p^n \phi_k/\p \tau^n|^2\rangle \sim (dL/d \tau)^{2n}L^{2-2n}h_n(kL),
\end{eqnarray}
where $h_n(kL)=k^2L^2 h_{n-1}(kL) \sim (kL)^{2n-3}$ as $k \gtrsim 1/L$. Therefore we have
\begin{eqnarray}
\frac{\p^n \phi_k}{\p \tau^n}
&\sim& \left(\frac{dL}{d \tau}\right)^n L^{1-n} (kL)^{n-3/2},
\nonumber \\
&\sim&
\cases{\tau^{-2n/3-1/6}k^{n-3/2} & \quad conserved
\cr \tau^{-n/2-1/4}k^{n-3/2} & \quad non-conserved \cr}
        \label{eq:fieldscaling}
\end{eqnarray}
The above expression is valid for $n \ge 0$ for two dimensional scalar
order parameter(s).

%%%%%%%%%%%%%%%%%%%%%%%%%%%%%%%%%%%%%%%%%%%%%%%%%%%%%%%%%%%%%%%%


\begin{thebibliography}{00}

\bibitem{Cahn1}J. W. Cahn and J. E. Hilliard, J. Chem. Phys. {\bf 28}, 258 (1958).
\bibitem{Allen1}S. M. Allen and J. W. Cahn, Acta Metall. {\bf 27}, 1085 (1979).
\bibitem{Wiltzius1} P. Wiltzius and A. Cumming, Phys. Rev. Lett. {\bf 66}, 3000 (1991).
\bibitem{Shannon1} R. F. Shannon, S. E. Nagler, C. R. Harkless and R. M. Nicklow,
Phys. Rev. B {\bf 46}, 40 (1992).
\bibitem{Gaulin1} B. D. Gaulin, S. Spooner and Y. Morii, Phys. Rev. Lett. {\bf 59}, 668 (1987).
\bibitem{Mason1} N. Mason, A. N. Pargellis and B. Yurke, Phys. Rev. Lett. {\bf 70}, 190 (1993).
\bibitem{Chuang1} I. Chuang, N. Turok and B. Yurke, Phys. Rev. Lett. {\bf 66}, 2472 (1991).
\bibitem{Laguna1} P. Laguna and W. H. Zurek, Phys. Rev. Lett. {\bf 78}, 2519 (1997).
\bibitem{Bray1}A. J. Bray, Adv. Phys. {\bf 43}, 357 (1994).
\bibitem{Oono1}Y. Oono and S. Puri, Phys. Rev. Lett. {\bf 58}, 836 (1987).
\bibitem{Oono2}Y. Oono and S. Puri, Phys. Rev. A {\bf 38}, 434 (1998).
\bibitem{Puri1}S. Puri and Y. Oono, Phys. Rev. A {\bf 38}, 1542 (1988).
\bibitem{Shinozaki1}A. Shinozaki and Y. Oono, Phys. Rev. E {\bf 48}, 2622 (1993).
\bibitem{Zapotocky1}M. Zapotocky, P. M. Goldbart and N. Goldenfeld,
Phys. Rev. E {\bf 51}, 1216 (1995).
\bibitem{Teixeira1}P. I. C. Teixeira and B. M. Mulder, Phys. Rev. E {\bf 55}, 3789 (1997).
\bibitem{Oono3}Y. Oono, Phys. Rev. E {\bf 55}, 3792 (1997).
\bibitem{Chen1}L. Q. Chen and J. Shen, Comput. Phys. Commun. {\bf 108}, 147 (1998).
\bibitem{Zhu1}J. Zhu, L. Q. Chen, J. Shen and V. Tikare, Phys. Rev. E {\bf 60}, 3564 (1999).
\bibitem{Eyre1}D. J. Eyre, in {\em Computational and Mathematical Models of
Microstructural Evolution}, edited by J. W. Bullard {\em et al.} (The Material
Research Society, Warrendale, PA, 1998), pp. 39-46.
\bibitem{Eyre2}D. J. Eyre, http://www.math.utah.edu/ $\tilde{}$ eyre/research/methods/stable.ps.
\bibitem{Vollmayr1}B. P. Vollmayr-Lee and A. D. Rutenberg, Phys. Rev. E {\bf 68}, 66703 (2003).
\bibitem{Cheng1}M. Cheng and A. D. Rutenberg, Phys. Rev. E {\bf 72}, 055701(R) (2005).
\bibitem{Cheng2}M. Cheng and J. A. Warren, Phys. Rev. E {\bf 75}, 017702 (2007).
\bibitem{Elder1}K. R. Elder, M. Katakowski, M. Haataja, and M. Grant, Phys. Rev. Lett. {\bf 88},
245701 (2002).
\bibitem{Elder2}K. R. Elder and M. Grant, Phys. Rev. E {\bf 70}, 051605 (2004).
\bibitem{Cheng3}M. Cheng and J. A. Warren, J. Comput. Phys. {\bf 227}, 6241 (2008).
\bibitem{Bray2}A. J. Bray and A. D. Rutenberg, Phys. Rev. E {\bf 49}, R27 (1994).
\bibitem{Rutenberg1} A. D. Rutenberg and A. J. Bray, Phys. Rev. E {\bf 51}, 5499 (1995).

%%%%%%%%%%%%%%%%%%%%%%%%%%%%%%%%%%%%%%%%%%%%%%%%%%%%%%%%%%%%%%%%%



% notes:

% \bibitem{label2}
% If there is a note, it should come last:
% \bibitem{label3} \note
% \end{subbibitems}

%\bibitem{}

\end{thebibliography}
\end{document}